\documentclass{ws-procs975x65}

\begin{document}



\title{Thick brane solution with two scalar fields}

\author{VLADIMIR DZHUNUSHALIEV
\footnote{Senior Associate of the Abdus Salam ICTP}}

\address{Dept. Phys. and Microel. 
Engineer., Kyrgyz-Russian Slavic University, Bishkek, Kievskaya Str. 
44, 720021, Kyrgyz Republic\\
\email{dzhun@krsu.edu.kg}}

\author{HANS-J\"URGEN SCHMIDT}

\address{Inst. Math., University Potsdam, Am Neuen Palais 10, D-14469
Potsdam, Germany\\
\email{hjschmi@rz.uni-potsdam.de}
}

\author{KAIRAT MYRZAKULOV}

\address{Institute of Physics and 
Technology, 050032, Almaty, Kazakhstan\\
\email{}}

\author{RATBAY MYRZAKULOV}

\address{Institute of Physics and 
Technology, 050032, Almaty, Kazakhstan\\
\email{cnlpmyra@mail.ru}}


\begin{abstract}
A new 5D thick brane solution is presented. We conjecture 
that the deduced thick brane is a plane defect in a bulk gauge condensate. 
\end{abstract}

\bodymatter

\section{Introduction}\label{intro}
In recent years there has been a revived interest in theories having a larger
number of  spatial dimensions than the three that are observed. In contrast 
to the original  Kaluza-Klein theories of extra dimensions, the recent development 
of extra dimensional  theories allow the extra dimensions to be large and even infinite 
in size (in the  original Kaluza-Klein theories the extra dimensions were 
curled up or compactified to the experimentally unobservable small size of the 
Planck length: $10^{-33}$ cm). These new  extra dimensional theories have opened 
up new avenues to explaining some of the open  questions in particle physics 
(the hierarchy problem, nature of the electro-weak symmetry  breaking, 
explanation of the family structure) and astrophysics (the nature of dark matter, the 
nature of dark energy). In addition they predict new experimentally measurable 
phenomena in high precision gravity experiments,  particle accelerators, and 
in astronomical observations. 
\par
\begin{figure}[h]
  \begin{center}
  \fbox{
  \includegraphics[height=5cm,width=5cm]{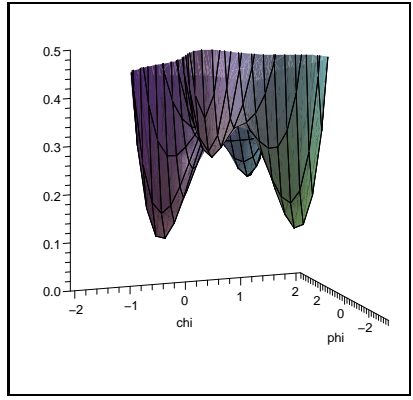}}
  \caption{The profile of the potential $V(\phi, \chi)$.}
  \label{fig3}
  \end{center}
\end{figure}

 
We consider 5D gravity + two interacting fields \cite{Dzhunushaliev:2006vv}. The 
key for the existence of a regular solution
here is that the scalar fields potential have to have \emph{local} 
and \emph{global} minima, and at infinity the scalar fields tend to a 
local but  \emph{not} to global minimum. The 5D metric is 
\begin{equation}
     ds^2 = a(y) \eta_{\mu \nu} dx^\mu dx^\nu - dy^2 .
\label{sec2-10}
\end{equation}
The Lagrangian for the scalar fields $\phi$ and $\chi$ is 
\begin{equation}
     \mathcal L = \frac{1}{2} \nabla_A \phi \nabla^A \phi + 
     \frac{1}{2} \nabla_A \chi \nabla^A \chi - V(\phi, \chi)     ,
\label{sec2-20}
\end{equation}
where $A= 0,1,2,3,5$. The potential $V(\phi, \chi)$ is 
\begin{equation}
     V(\phi, \chi)  = \frac{\lambda_1}{4} \left(
          \phi^2 - m_1^2
     \right)^2 + 
     \frac{\lambda_2}{4} \left(
          \chi^2 - m_2^2
     \right)^2 + \phi^2 \chi^2 - V_0 ,
\label{sec2-30}
\end{equation}
where $V_0$ is a constant which can be considered as a 5D cosmological 
constant $\Lambda$. The profile of the potential $V(\phi, \chi)$ is presented 
in Fig. \ref{fig3}. 
The corresponding field equations are 
\begin{eqnarray}
     \frac{a''}{a} - \frac{a'^2}{a^2} &=& 
          - \frac{1}{2}\left( \phi'^2 + \chi'^2 \right),
\label{sec2-75}\\
     \frac{a'^2}{a^2} &=& \frac{1}{8} \left[
          \phi'^2 + \chi'^2 - \frac{\lambda_1}{2} \left( 
               \phi^2 - m_1^2
          \right)^2 - \frac{\lambda_2}{2} \left( 
               \chi^2 - m_2^2
          \right)^2 - 
          \phi^2 \chi^2 + 2 V_0 
     \right] ,
\label{sec2-85}\\
     \phi'' + 4 \frac{a'}{a} \phi' &=& \phi \left[
          \chi^2 + \lambda_1 \left( \phi^2 - m_1^2 \right)
     \right] ,
\label{sec2-95}\\
     \chi'' + 4 \frac{a'}{a} \chi' &=& \chi \left[
          \phi^2 + \lambda_2 \left( \chi^2 - m_2^2 \right)
     \right] .
\label{sec2-105}
\end{eqnarray}
The boundary conditions are 
\begin{eqnarray}
     a(0) &=& a_0 , \; a'(0) = 0 ,\;
     \phi(0) = \phi_0 , \; \phi'(0) = 0 , \;
     \chi(0) =\chi_0 , \; \chi'(0) = 0 . 
\label{sec2-140}
\end{eqnarray}
The boundary condition \eqref{sec2-140} and Eq. \eqref{sec2-85} 
give us the following constraint 
\begin{equation}
     V_0 = \frac{\lambda_1}{4} \left( 
               \phi^2_0 - m_1^2
          \right)^2 + \frac{\lambda_2}{4} \left( 
               \chi^2_0 - m_2^2
          \right)^2 + \frac{1}{2} \phi^2_0 \chi^2_0,
\label{sec2-150}
\end{equation}
This means that the constant $V_0$ is not an arbitrary constant but it is defined by the
$Z_2$--symmetry of the thick brane. 
\par
\begin{figure}[h]
\begin{minipage}[t]{.45\linewidth}
  \begin{center}
  \fbox{
  \includegraphics[height=5cm,width=5cm]{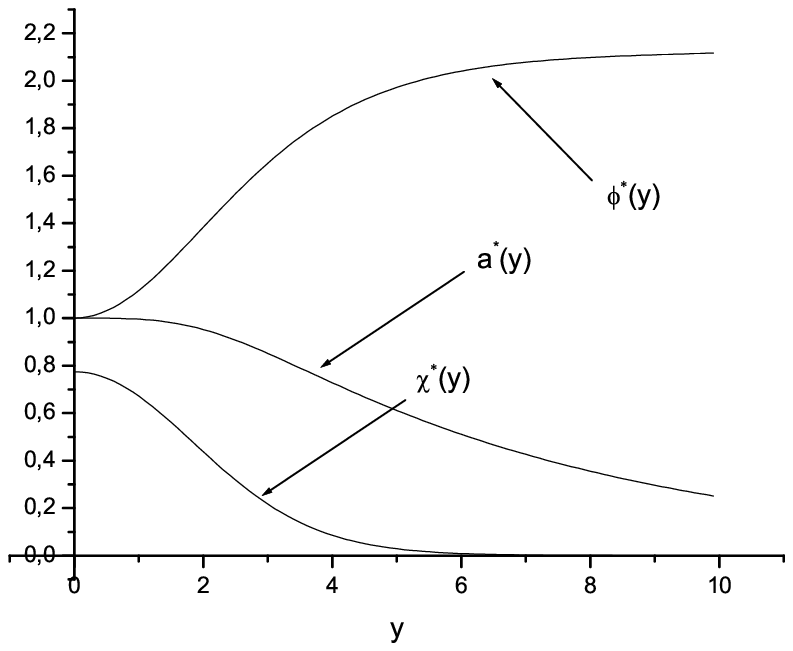}}
  \caption{The functions $a^*(y), \phi^*(y), \chi^*(y)$}
  \label{fig1}
  \end{center}
\end{minipage}\hfill
\begin{minipage}[t]{.45\linewidth}
  \begin{center}
  \fbox{
  \includegraphics[height=5cm,width=5cm]{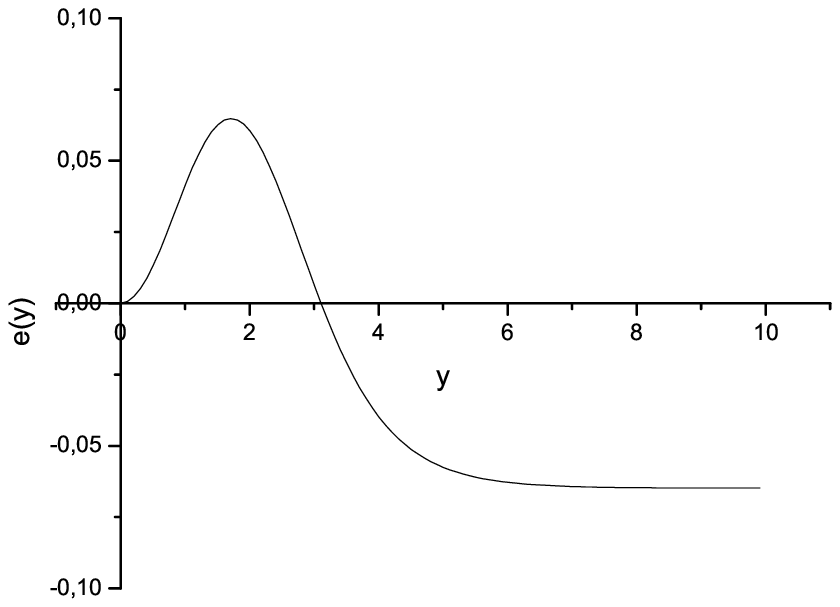}}
  \caption{The profile of dimensionless energy density.}
  \label{fig2}
  \end{center}
\end{minipage}\hfill 
\end{figure}

The mathematical problem for solving  these equations  is that \emph{not for 
all} values of  the parameters $m_{1,2}$ \emph{a regular solution exists.} Thus the 
problem of finding a regular solution of these equations  is \emph{a non-linear  
eigenvalue problem} for the parameters \emph{$m_{1,2}$} and the 
eigenfunctions \emph{$\phi, \chi$}. In Fig. \ref{fig1} we present \emph{the regular solution} which describes the thick brane solution for the 5D gravity. The dimensionless energy density is presented in Fig. \ref{fig2}. The thickness of the brane depends on all parameters 
which are included in the equations: $\lambda_{1,2}, \phi(0), \chi(0).$

\section{Main features of the presented solution}

\begin{itemize}
     \item At the infinity the scalar fields tends to \emph{a local minimum not to global}.
     \item The first item leads to the fact that the solution is \emph{topologically trivial}.
     \item Mathematically the solution of Einstein-scalar fields equations is \emph{a
non-linear eigenvalue problem.} 
     \item There are arguments (\cite{d2})  that these scalar 
fields present non-perturbatively quantized SU(3) gauge field. In this case the brane 
world is a plain defect in 5D spacetime filled with a gauge condensate.
\end{itemize}

More background material can be found in Refs. \cite{d3} and \cite{d4}.

\vfill
\end{document}